# Ultrathin van der Waals metalenses

Chang-Hua Liu[1,2,*], Jiajiu Zheng[3,+], Shane Colburn[3,+], Taylor K. Fryett[3], Yueyang Chen[3], Xiaodong Xu[1,4], Arka Majumdar[1,3, *]

[1] Department of Physics, University of Washington, Seattle, WA, USA.

[2] Institute of Photonics Technologies, National Tsing Hua University, Hsinchu 30013, Taiwan

[3] Department of Electrical Engineering, University of Washington, Seattle, WA, USA.

[4] Department of Materials Science and Engineering, University of Washington, Seattle, WA, USA

[*] *Corresponding Author: chliu@ee.nthu.edu.tw; arka@uw.edu*

[+] *These two authors contributed equally.*

**Ultrathin and flat optical lenses are essential for modern optical imaging, spectroscopy, and energy harvesting[1-5]. Dielectric metasurfaces comprising nanoscale quasi-periodic resonator arrays are promising for such applications, as they can tailor the phase, amplitude, and polarization of light at subwavelength resolution, enabling multi-functional optical elements[2-6]. To achieve $2\pi$ phase coverage, however, most dielectric metalenses need a thickness comparable to the wavelength, requiring fabrication of high-aspect-ratio scattering elements. Here, we report ultrathin dielectric metalenses made of van der Waals (vdW) materials, leveraging their high refractive indices and the incomplete phase design approach[7,8] to achieve device thicknesses down to $\sim\lambda/10$, operating at infrared and visible wavelengths. These materials have generated strong interest in recent years due to their advantageous optoelectronic properties[9-13]. Using vdW metalenses, we demonstrate near diffraction-limited focusing and imaging, and exploit their layered nature to transfer the fabricated metalenses onto flexible substrates to show strain-induced tunable focusing. Our work enables further downscaling of optical elements and opportunities for integration of metasurface optics in ultra-miniature optoelectronic systems.**

Sub-wavelength diffractive optical elements, also known as metasurfaces, have fostered considerable interest in the photonics community in recent years. A sub-wavelength scatterer can provide different phase-shifts to the incident light by virtue of their lateral geometry, while having the same thickness. A metasurface exploits this effect to realize ultra-thin and flat optical elements, with uniform thickness, which can be fabricated by a single stage lithography. Although several works in this field have successfully demonstrated metalenses[14,15] and other optical components[2,3,16] using deep sub-wavelength-thick metallic nano-antennas, these devices usually suffer from high absorption loss and strong polarization sensitivity. An alternative approach is to exploit arrays of dielectric resonators[17,18]. Thus far, a diverse set of low-loss dielectric metasurface optical elements operating in the infrared and visible wavelength ranges have been reported[4,17-20]. However, these dielectric elements are much thicker (on the order of the wavelength) to achieve a full $2\pi$ phase shift. At the same time, to impart small phase shifts the resonators need to have narrower lateral dimensions. As the thickness is uniform for all the scatterers, several pillars end up having high high-aspect-ratio (typically > 5). Fabricating these delicate structures while simultaneously keeping a low sidewall roughness is a challenging nanofabrication task and may limit the practical applications of metalenses. Although thinner and efficient dielectric metasurfaces can be ideally realized at near-infrared wavelengths using high-index materials with low absorption loss, such as group III-V compounds like AlGaAs (n ~3.8) or GaP (n ~3.4), it is challenging to directly grow these materials on different substrates.

Layered van der Waals (vdW) materials exhibit diverse optoelectronic properties, ranging from wide band-gap hexagonal boron nitride (h-BN) and excitonic transition metal dichalcogenides (TMDCs), to semi-metallic graphene[10,11]. These materials can be transferred on any substrate without requiring explicit lattice matching. This presents new opportunities for creating hybrid nano-structures that can simultaneously take advantage of industrial semiconductor manufacturing technologies, and benefit from the unique properties of vdW materials[9,11,21]. In recent years, these materials have been used in conjunction with integrated nano-photonic structures to demonstrate various active devices, for short-distance optical communication and energy harvesting applications[10,12,13,22,23]. Recently, nano-patterned periodic sub-wavelength gratings made of vdW materials at mid-infrared frequencies have been reported[24], although wavefront engineering in the far-field has not been demonstrated. Additionally, ultrathin Fresnel lenses made of molybdenum disulfide have been demonstrated[25]. These lenses, however, do not exploit nano-meter scale patterning and rely on phase accumulation from the variable thickness. Varying the thickness in this manner necessitates use of low-throughput fabrication technology, such as focused ion-beam milling.

In this paper, we design and fabricate the first vdW metalenses, with significantly reduced thickness ($\sim 0.1\lambda - 0.5\lambda$) compared to the optical wavelength by exploiting the incomplete phase-based design approach[7,8] and the high refractive indices of vdW materials. We validate our design via fabrication and characterization of metalenses at visible and near-infrared wavelengths. Due to the nature of vdW interactions, our proposed metalenses can be readily transferred onto diverse material platforms (Fig. 1a), distinct from typical dielectric metasurfaces, limited by bottom-up material growth.

To design the metalenses, we first compute the transmission properties of periodic arrays of cylindrical vdW-nanopillars using rigorous coupled-wave analysis (RCWA)[26]. The pillars are made of wide band-gap h-BN ($E_g \sim 5.2$ eV, refractive index $n \sim 2.17$)[27] and rest on a quartz substrate with a lattice periodicity of $p \sim$ 260 nm (Fig. 1b). The wavelength of incident light is $\lambda = 450$ nm. By varying the dimensions of the pillars (the thickness $t$ and diameter $D$), the transmitted light's phase and amplitude can be tuned (Fig. 1c-e). This optical response can be explained by considering each pillar as a low-quality-factor Fabry-Perot resonator. The propagation phase increases with the dimensions of the resonator and varies rapidly near the supported resonant modes. Based on the selected dielectrics and designed parameters, the pillars should be thicker than 450 nm to achieve a complete $2\pi$ phase control.

To reduce the thickness of the scatterers while mitigating their drop in performance, we adapt the theory of incomplete phase modulation, originally developed for creating efficient liquid crystal Fresnel optics[7,8]. For a spherical singlet lens, the addressed phase ($\varphi$) profile should follow:

$$\varphi(r) = -\frac{\pi r^2}{\lambda F} \quad (1)$$

where $r$, $\lambda$, and $F$ are the radial coordinate, optical wavelength, and focal length of the designed lens, respectively. With limited phase coverage, i.e., the maximum phase shift being less than $2\pi$ (Fig. 2a), the implemented phase functions can be correlated with the addressed 0-$2\pi$ phase ($\varphi$) via a saturated mismatched model (Figs. 2b, c, green curves). This model is derived from the minimum Euclidean distance principle, and should lead to higher focusing efficiency[8,28] compared to a simple linear interpolation (Figs. 2b, c, blue curves, see Methods for details).

To experimentally demonstrate this concept, a 245-nm-thick ($\sim 0.54 \lambda$) exfoliated h-BN flake was transferred onto a quartz substrate and then patterned into an array of pillars to form four different metalenses with a focal length of $F = 160$ μm at $\lambda = 450$ nm (see Methods for nanofabrication) and a lens diameter of 60 μm. Among them, lenses A and C comprise pillars that can only implement phase shifts from 0 to $0.7\pi$, while lenses B and D have a more limited phase modulation depth, ranging from 0 to $0.5\pi$

(Fig. 2d). The metalenses C and D are designed using linear interpolation (following Figs. 2b, c, blue curves), while A and B are developed with the saturated mismatched model (following Figs. 2b, c, green curves). We measure the focusing performance of the metalenses using the experimental setup schematically illustrated in Fig. S1 (see Supplementary Section 1). Figures 2e-f show the intensity profile measured in the designed focal plane. Despite the limited range of the implemented phase, all four lenses exhibit focusing. The corresponding intensity cuts through the center of their focal spots show that the full-width at half-maximum (FWHM) for all the metalenses is ~2.8 μm, close to the diffraction-limited FWHM of 2.3 μm. Moreover, the focusing from metalenses B and C indicates that metalenses developed with the saturated mismatched model (lens B) with smaller phase modulation can have higher focal intensity than a lens (lens C) developed using the linear interpolation model even with a larger phase modulation depth (Fig. 2f). Specifically, we found that in using the saturated mismatched model, the focusing efficiency could be increased by ~1.5 times while having the same phase modulation. We note these results not only agree with our numerical simulations, but also indicate the possibility of creating thin and efficient dielectric lenses with limited phase coverage (See Supplementary Section 2 and 3 for further discussions). Thus, with this design principle, we can relax the stringent requirements on the scatterers to reach a full $2\pi$ phase shift. This in turn avoids the necessity of fabricating high-aspect-ratio structures and provides better tolerance to fabrication and simulation errors.

To extend the metalenses to a longer wavelength regime, we use TMDC, specifically molybdenum disulfide ($MoS_2$), to make metalenses. $MoS_2$ has a high refractive index[29] that in combination with incomplete phase design can potentially push the thickness of dielectric metalenses to far below the operating wavelength. To verify this concept, we fabricated a metalens based on 190-nm-thick (~0.14 $\lambda$) $MoS_2$ pillars with a designed focal length of 300 μm and a lens diameter of 100 μm (See methods for nanofabrication). The simulated optical response of arrays of $MoS_2$ pillars at $\lambda = 1310$ nm is shown in Figure 2g, and the phase profile of the lens was developed using the saturated mismatch model with the implemented phase shift ranging from 0 to $0.7\pi$. By using the characterization setup shown in Figure S1, we can examine the intensity profile of this designed lens and resolve a near diffraction-limited focal spot (FWHM~5 μm) at $F = 300$ μm (Figures 2h-i), which agrees with our design parameters, showing the promise for achieving a compact infrared imaging system.

With the success of our as-fabricated devices for focusing, we next explore the integrability of vdW metalenses in systems. In this experiment, a 120-nm-thick exfoliated h-BN flake was patterned into a metalens with focal length $F = 250$ μm on a quartz substrate (Supplementary Section 8). This pre-patterned device was subsequently transferred onto a stretchable polydimethylsiloxane (PDMS) substrate using the dry transfer pick-up technique[11,21]. To characterize the optical properties and tunability of the transferred device, we resolve the transmitted light intensity profiles along the axial plane as a function of radial strain applied to the PDMS (Fig. 3a). In the absence of strain, the focal length of the lens is at the designed 250 μm, suggesting the metalens remains intact after the transfer procedure. In addition to the focal spot at $F$, multiple high-intensity spots can be observed at $\frac{F}{N}$, where $N$ is an integer. This behavior originates from the incomplete phase implementation (Supplementary Section 4 to 5). In applying the radial strain ($\epsilon$), the measured focal spot shifts away from the initial focal plane of the metalens (Supplementary Section 6). This tuning behavior is linearly proportional to $(1+\epsilon)^2$ (Fig. 3b), which not only agrees with the theoretical prediction[30,31], but also highlights the opportunities of utilizing vdW materials for making tunable metasurfaces.

Finally, we examine the efficacy of the design and fabrication process by capturing images with the h-BN metalenses in Fig. 2d using a custom-built imaging system (Fig. S1). The scattered light from a printed Mona Lisa pattern is captured using metalenses A, B and C, which were imaged by a CMOS camera (Fig.

4a). Critically, these results demonstrate that our proof-of-concept devices can form images. Metalens A, designed via the saturated mismatched model, exhibits superior performance relative to the other two, evidenced by its greater luminance and sharpness. This is consistent with the fact that the saturated mismatched model is more efficient compared to the linear mismatched model, which generates a lower intensity image. We note this imaging quality can be further improved by using computational post-processing[32] (Fig. 4b, Supplementary Section 7) or by increasing the numerical aperture of the metalenses (Fig. S7), making our developed thin metalenses more attractive for imaging applications.

In summary, we demonstrated a viable route to realize ultrathin dielectric metalenses based on vdW materials. This adapted design principle can be applied to different spectral regions, a broad class of dielectric materials, and circumvents the current fabrication challenge of making high-aspect-ratio nanoscale scatters. Additionally, we show that utilizing vdW materials would offer a unique opportunity to realize metalenses which are integrable with other substrates, including flexible polymers. It is also noteworthy that many emerging layered materials exhibit diverse optical properties, such as phase-change or nonlinear behaviors, and such materials could be further stacked together to form more complicated vdW heterostructures with the ability to tailor material properties with atomic precision.[11] A tremendous potential exists for creating vdW metasurfaces with novel functionalities for applications in optical sensing, imaging, focal length tuning, and energy harvesting.

**Figure captions:**

**Figure 1 Process of transferring vdW materials and numerical simulations. a.** The schematic diagram shows the nanostructured vdW materials can be transferred onto different substrates using a polycarbonate (PC) film. **b.** Schematic illustration of arrays of vdW nanopillars, composed of h-BN and spaced apart by $p = 260$ nm. Varying the diameter ($D$) and thickness ($t$) of the pillars affects the transmission properties of incident light. **c-d.** The color maps show RCWA providing the change in the (**c**) amplitude and (**d**) phase of a wavefront after passing through the array when varying the dimensions of the nanopillars. **e.** Left to right: Pillar diameter versus the change in amplitude (black) and phase (red) of a transmitted wavefront with the thickness of the nanopillars fixed at 300, 400, and 500 nm respectively. These simulated results are extracted from (**c**) and (**d**).

**Figure 2 Design principle and characterizations of vdW metalenses. a.** Simulated change in the amplitude (black) and phase (red) of a wavefront passing through a 245-nm-thick hBN pillar array. The operating wavelength is 450 nm and the refractive index of hBN is ~2.17. **b-c.** The implemented phase versus the addressed phase are correlated with linear (blue) and saturated (green) mismatched models. The range of implemented phase is (b) 0-0.5π (diameters ranging from 117 nm to 195 nm), and (c) 0-0.7π (diameters ranging from 90 nm to 210 nm). **d.** Optical image of the fabricated metalenses. Device A: 0.7π phase modulation and the saturated mismatched model; Device B: 0.5π phase modulation and the saturated mismatched model; Device C: 0.7π phase modulation and the linear mismatched model; Device D: 0.5π phase modulation and the linear mismatched model. Scale bar: 30 μm. **e.** Intensity distributions in the focal plane of four metalenses. **f.** Line cuts of the intensity distributions along the four focal spots shown in (**e**). Their field profiles are fitted to a Gaussian function. The data is normalized with respect to the peak intensity of Device A. **g.** Simulated change in the amplitude (black) and phase (red) of a wavefront passing through an array of 190-nm-thick $MoS_2$ pillars with a periodicity of 540 nm. The operating wavelength is 1310 nm and the refractive index of $MoS_2$ is ~4.2 [29]. Inset: Scanning electron micrograph (SEM) of the portion of ultrathin $MoS_2$ pillars. The sample was covered with 5-nm-thick gold to minimize charging. Scale bar: 500 nm. **h.** Intensity profile measured along the axial plane of the $MoS_2$ metalens with an incident wavelength of 1310 nm. **i.** Normalized intensity profile measured along the focal plane at 300 μm in (**h**) and its corresponding Gaussian fit.

**Figure 3 Integrable and tunable metalenses based on vdW materials. a.** Three representative intensity profiles measured along the axial plane of h-BN metalens integrated onto a PDMS substrate. Left to right: The strain applied to the PDMS substrate are 0%, 9.5%, and 25.6%, respectively. **b.** Measured (round symbols) and analytically predicted (solid line) focal lengths under different strain values.

**Figure 4 Vdw metalenses for optical imaging application. a.** Images formed by metalens A (bottom), B (top left), and C (top right), simultaneously captured using a CMOS imaging sensor. **b.** The same images shown in (**a**) after computational post-processing.

**Methods:**

**Linear and saturated mismatched models.** The two models describe the relations between the implemented phase (*I*), used for building the metalens, versus the addressed 0-2π phase (*φ*). Given the implemented phase is limited between aπ-bπ (0<*a*< *b* <2), the linear mismatched model correlates the two phases via the relation:

$$I(\varphi) = a\pi + \frac{(b-a)\times\varphi}{2},$$

while for the saturated mismatched model, the relation is as below:

$$I(\varphi) = a\pi + \varphi, \text{if } \varphi < (b-a)\pi$$

$$I(\varphi) = b\pi, \text{if } (b-a)\pi \leq \varphi < \frac{(b-a)\pi}{2} + \pi$$

$$I(\varphi) = a\pi, \text{if } \frac{(b-a)\pi}{2} + \pi \leq \varphi$$

**Device fabrication.** We mechanically exfoliated vdW materials onto $SiO_2$/Si substrates and used atomic force microscopy (AFM) to identify their thickness and cleanliness. The selected flakes were transferred from $SiO_2$/Si onto quartz substrates using the dry transfer pick-up technique[11,21], and then covered by a 50-nm-thick aluminum film, blanket-deposited by radiofrequency sputtering. This aluminum thin film was applied as a dry-etching mask and charge dissipation layer for e-beam lithography. The flakes were subsequently patterned into nanopillars using e-beam lithography and dry etching processes. Finally, the e-beam resist and aluminum film were removed by acetone and AD-10 photoresist developer respectively to form the dielectric metalenses.

**Acknowledgement:** We acknowledge Mr. Alan Zhan for input on fabrication. This work is supported by the AFOSR grant FA9550-18-1-0104 (program manager Dr. Gernot Pomrenke) and NSF MRSEC 1719797. All the fabrication processes were performed at the Washington Nanofabrication Facility (WNF), a National Nanotechnology Infrastructure Network (NNIN) site at the University of Washington, which is supported in part by the National Science Foundation (awards 0335765 and 1337840), the Washington Research Foundation, the M. J. Murdock Charitable Trust, GCE Market, Class One Technologies and Google.

**Author Contributions:**
C.H.L. and A.M. conceived the idea. C.H.L. and S.C. performed the design and simulation. J.Z. fabricated all the devices. T.K.F. and Y.C. helped with material search and AFM characterization. C.H.L. performed the measurement and data analysis with help from S.C. X.X. and A.M. supervised the whole project.

**Competing Interests:**

The authors declare no competing interests.

**Figure 1**

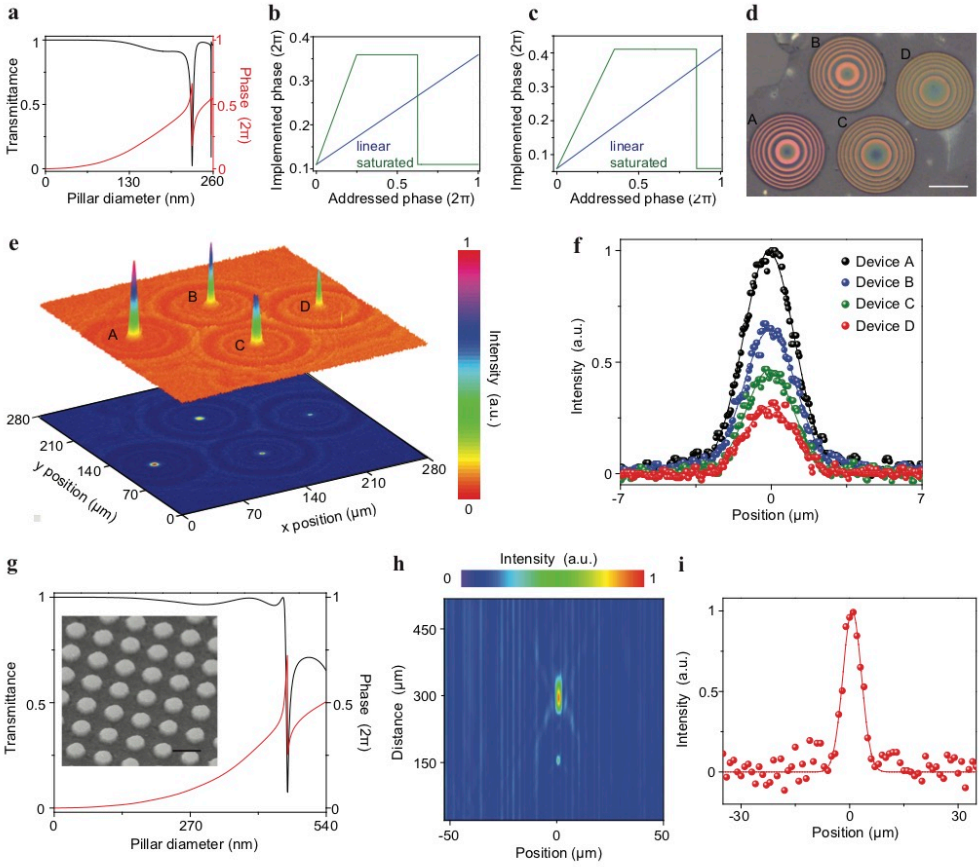

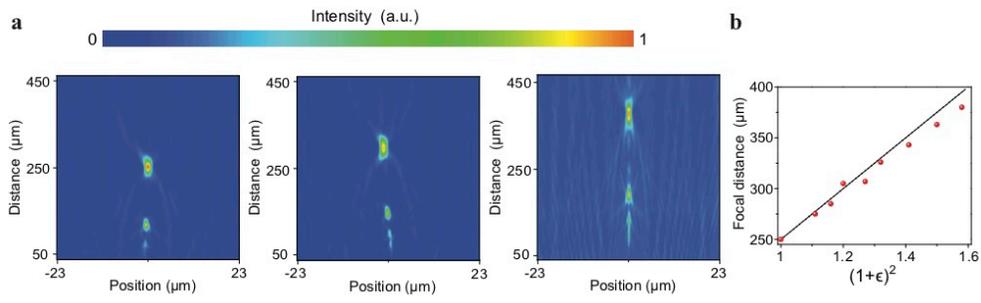

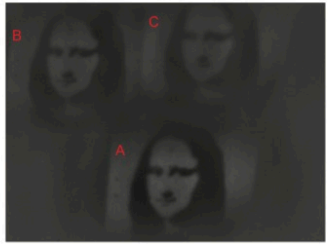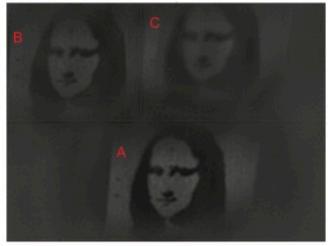

# Ultrathin van der Waals metalenses

Chang-Hua Liu[1,2,*], Jiajiu Zheng[3,+], Shane Colburn[3,+], Taylor K. Fryett[3], Yueyang Chen[3], Xiaodong Xu[1,4], Arka Majumdar[1,3,*]

1. Department of Physics, University of Washington, Seattle, WA, USA.

2. Institute of Photonics Technologies, National Tsing Hua University, Hsinchu 30013, Taiwan

3. Department of Electrical Engineering, University of Washington, Seattle, WA, USA.

4. Department of Materials Science and Engineering, University of Washington, Seattle, WA, USA

[*] *Corresponding Author:* chliu@ee.nthu.edu.tw; arka@uw.edu

[+] *These two authors contributed equally.*

**Contents:**

## Supplementary Section 1.

## Measurement setup and procedure

The measurement setup schematically shown in Fig. S1 was built to characterize the focal spots, field profiles, and images produced by the fabricated metalenses. The designed metalenses in this work were used to collect the light emitted from fiber-coupled light emitting diodes or scattered from printed object patterns for imaging experiments. For the imaging experiment shown in Fig. 4a (main text), the imaged object is a black and white picture of the Mona Lisa printed on standard white printer paper. A custom-built microscope comprising an objective (Nikon Plan Fluor, NA = 0.75, WD = 0.66 mm) and tube lens (Thorlabs ITL200) magnified and projected the focal spot or image onto a camera (AmScope MU300 and Xenics Bobcat-1.7-320 for visible and infrared characterization respectively). The objective was mounted on a motorized translation stage so that this setup could resolve the field profiles over a wide range of depths along the optical axis of the metalenses. The captured images were analyzed and deconvolved using custom MATLAB scripts.

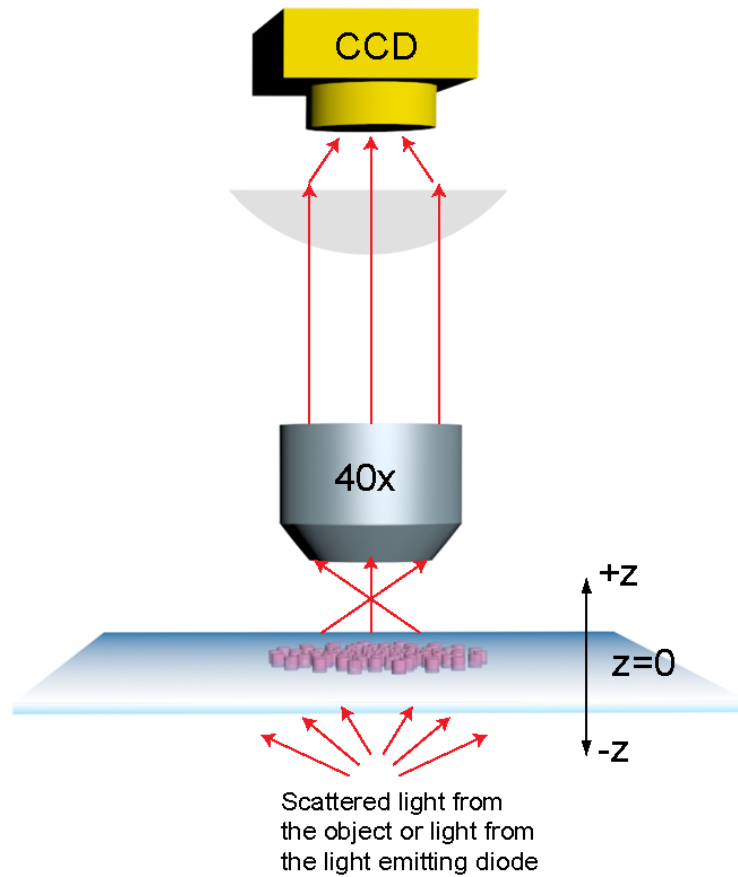

**Figure S1:** Schematic illustration of the measurement setup for characterizing the metalenses (not to scale).

**Supplementary Section 2.**

**Theoretical focusing efficiency of a metalens developed by incomplete phase design**

For a typical singlet metalens, its phase profile ($\varphi$) should follow:

$$\varphi(r) = -\frac{\pi r^2}{\lambda F} \quad (1)$$

where $r$ is the radial coordinate, $\lambda$ is wavelength, and $F$ is the focal length of the designed lens. Here, $\varphi$ varies from 0-$2\pi$ and increases with $r$, as discussed in the main text. If a phase range less to $2\pi$ is achievable with the designed phase shifters, then the implemented $I[\varphi(r)]$ phase is less than $2\pi$; however, even with this limitation it is still possible to design an efficient metalens with its implemented phase related to the addressed phase ($\varphi(r)$) using: (1) linear and (2) saturated mismatched models.[1,2]

Here, we use the transfer function $t$ to represent the effect of these models. Thus, the implemented phase of the metalens can be written as:

$$I[\varphi(r)] = t[\varphi(r)] \quad (2)$$

As such, the phase delay $P[\varphi(r)]$ imparted by the metalens would be:

$$P[\varphi(r)] = \exp\{-it[\varphi(r)]\} \quad (3)$$

As $P(\varphi)$ is a periodic function (with period $2\pi$), equation (3) can be expanded into Fourier series:

$$\exp\{-it[\varphi(r)]\} = \sum_{n=-\infty}^{n=\infty} C_n \exp\left[-i\frac{\pi r^2}{\lambda(F/n)}\right] \quad (4)$$

where $n$ is an integer and the expansion coefficient $C_n$ has the form:

$$C_n = \frac{1}{2\pi}\int_0^{2\pi} \exp[it(\varphi)]\exp(-in\varphi)d\varphi \quad (5)$$

According to equation (4), it is clear that the lens has multiple focal distances at $F/n$, relative to the plane of the metalens, where $n$ is either a positive or negative integer. Focal distances are positive for converging lenses and negative for diverging lenses. For the lens with its designed focal length at $F$, the theoretical focusing efficiency is proportional to $|C_1|^2$, and here, we simulate the relation between the range of focusing efficiency of a metalens versus implemented phase. We do this for the transfer functions described by both the linear and saturated mismatch models. The simulation result shown in Fig. S2 indicates the lens developed using the saturated mismatched model provides higher efficiency, which agrees with our experimental results shown in Fig. 2 of the main text.

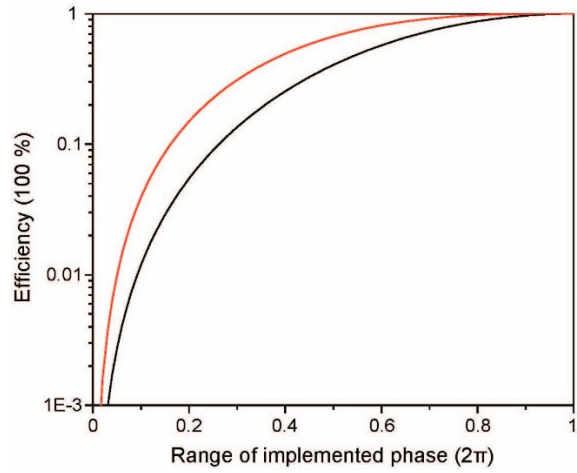

**Figure S2 Focusing efficiency versus the range of implemented phase:** The calculated result shows the theoretical limit of focusing efficiency of a metalens developed using the linear (black) and saturated mismatched (red) models, respectively.

**Supplementary Section 3.**

**Comparison of theoretical and experimental focusing efficiency**

To provide quantitative insight into the relative performance of the different incomplete phase design methods, we measured their efficiencies and also calculated their performance via the model in Supplementary Section 2. To measure the efficiency, we took the ratio of the power measured by integrating over the focal plane to that measured incident on the metasurface (Fig. 2e-f, main text). To calculate the theoretical efficiency, we used the model from Supplementary Section 2 to find the focusing efficiency. Table S1 details the relative performance of these lenses, taking the efficiency of Devices B-D and dividing by the efficiency of Device A, the lens designed used the saturated mismatched model. As detailed in table S1, the experimental data are in strong agreement with the theoretical values and demonstrate the superior performance of Device A.

|  | Theoretical prediction | Experimental result |
|---|---|---|
| Device B/Device A | 0.635 | 0.56 |
| Device C/Device A | 0.443 | 0.475 |
| Device D/Device A | 0.298 | 0.225 |

**Table S1 Ratio analysis of focusing efficiency of boron nitride metalenses.**

**Supplementary Section 4**

**Measuring the field profiles of a metalens developed by incomplete phase design**

To gain further insight into our developed metalens, we mounted the objective on a motorized translation stage to map the field profile of the device. The measurement results shown in Fig. S3a-b reveal that the metalenses developed by the incomplete phase design would converge (diverge) the light into (from) the points at $F/n$ relative to the plane of the metalens. We note this is in good agreement with the theoretical model, as discussed in supplementary section 2.

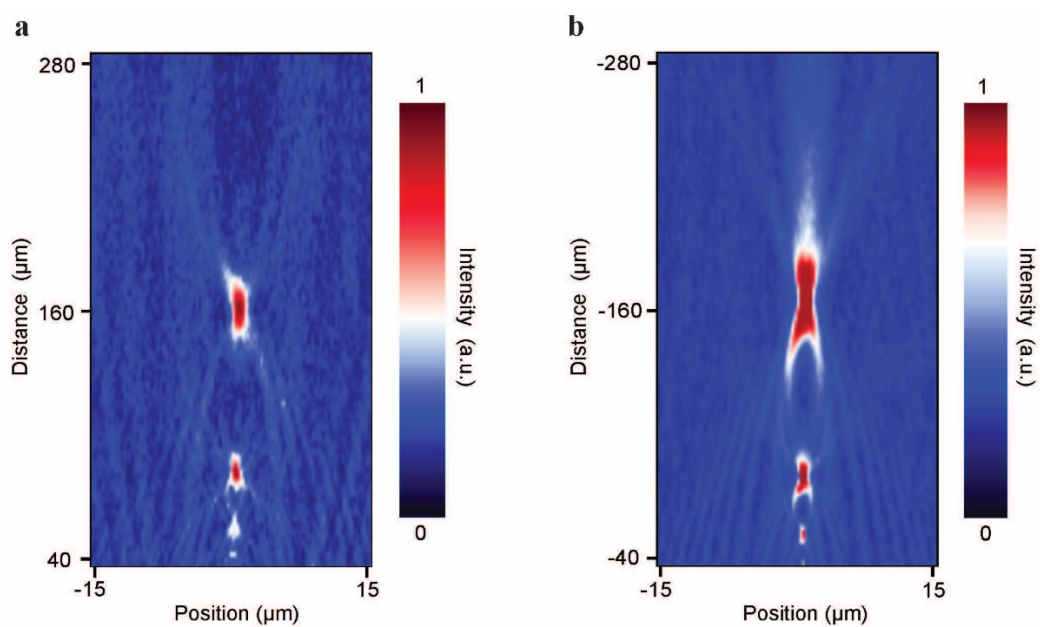

**Figure S3 Field profiles of a metalens developed using an incomplete phase design. a-b.** The filed profiles of device C shown in the main text are resolved by moving the focal plane of the 40x objective to **(a)** positive and **(b)** negative Z directions, respectively. Z=0 is the plane of the metasurface and the corresponding coordinates are depicted in Fig. S1.

**Supplementary Section 5.**

**Numerical analysis of the field profile of a metalens developed by incomplete phase design**

In addition to measuring the field profiles of our metalenses developed by incomplete phase design, we also simulated the designs to match with the theory. For this simulation, we model the propagation of light passing through the boron nitride metalens with 0.5 π phase modulation depth, with the implemented phase versus the addressed phase correlated via the linear mismatched model. To model this propagation, we treat the metalenses as complex amplitude masks, assigning the transmission coefficient calculated by rigorous coupled-wave analysis for a nanopost design to positions where the nanopost is placed. We then solve the Rayleigh-Sommerfeld diffraction integral by means of an angular spectrum propagator and propagate the electric field distribution to planes along the optical axis. As shown in Fig. S4, the simulated field profile exhibits multiple high intensity spots at $\frac{F}{N}$, in addition to the focal spot at $F$. This agrees with the analytical model (section 2) and our experimental results (section 4).

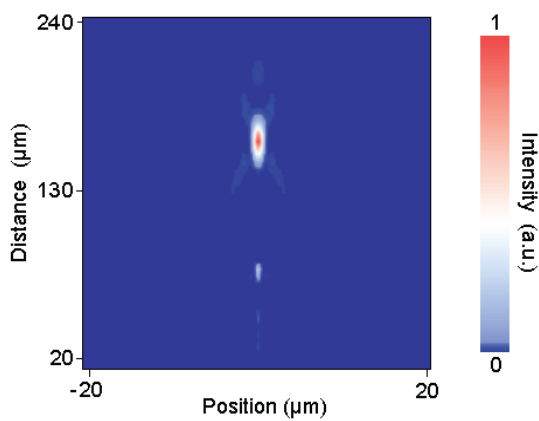

**Figure S4 Numerical simulation of the field profile of a metalens developed using an incomplete phase design.**

## Supplementary Section 6.

### Measurements of stretchable metalens

For stretchable measurements, the vdW metalens was transferred onto the square flexible (polydimethylsiloxane) PDMS substrate. The substrate was held by four clamps along each side. As applying tensile force to the clamps, the substrate would experience strain, causing the change of periodicity of nanopillars. To characterize the focal length, we first place the metalens at the focal plane of 40x objective, and then gradually move the objective towards +z direction, as shown in Fig. S1, to construct the intensity profiles along the axial plane of metalens. The same procedure was followed for each applied strain.

## Supplementary Section 7.

### Improving the quality of optical imaging by computational postprocessing[3]

To sharpen and improve the quality of the captured images, we applied a post-capture filter to partially deblur the images. Here, we use a linear Wiener filter with a signal-to-noise ratio (SNR) of 20. The array of Mona Lisa images (Fig. 4a, main text), where each image corresponds to that produced by a distinct lens, is segmented into three separate sub-images, which are locally deconvolved using the measured point spread functions for each lens (Fig. S5) and stitched back together to produce a final sharpened result (Fig. 4b, main text). The SNR was determined by hand tuning, adjusting the value until the images were the most aesthetically pleasing, balancing image sharpness with the tradeoff of noise amplification.

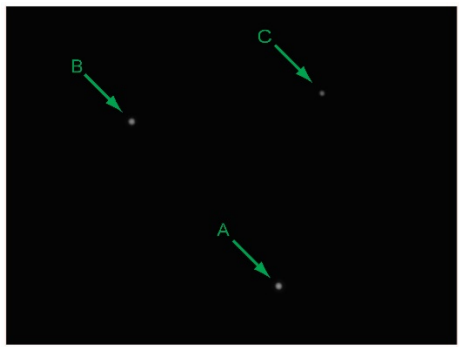

**Figure S5 Measured point spread functions.** This image was captured with a CMOS camera and it shows the focal spots of Device A, B, and C, respectively. This image was used for deconvolving the captured images created by these three separate metalenses.

## Supplementary Section 8.

### Design parameters of integrable and stretchable boron nitride metalens

The data of Fig. 3 in the main text is based on an exfoliated 120-nm-thick boron nitride flake, which was patterned into an array of varying diameter nanopillars (Fig. S6a). The designed pillars were simulated via rigorous coupled-wave analysis (RCWA) at λ=450 nm and their transmission properties are shown in Fig. S7b. To form the metalens with a focal length $F = 250$ μm, we utilized the pillars that could lead to phase shifts from 0 to $0.5\pi$ into the design, and these phase shifts are correlated with the addressed 0-$2\pi$ phase ($\varphi$) via the linear mismatched model.

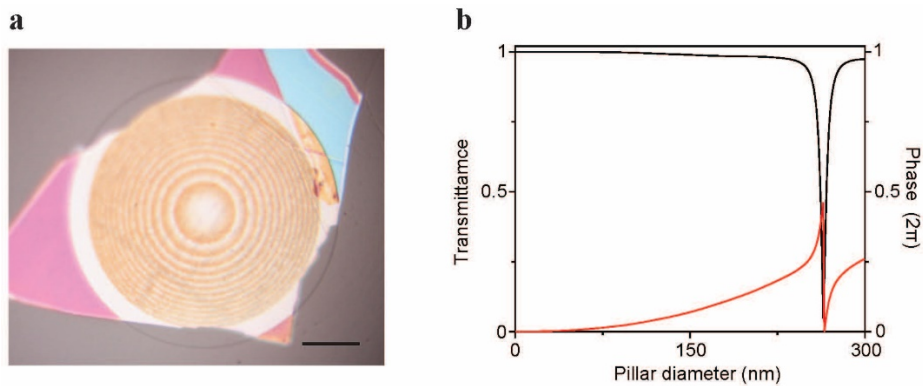

**Figure S6 Boron nitride metalens. a.** Optical image of the boron nitride metalens. Scalar bar, 20 μm **b.** RCWA analysis of pillar diameter versus the change in amplitude (black) and phase (red) of a transmitted wavefront. The spacing between pillars is fixed at 300 nm.

**Supplementary Section 9.**

**Increasing the diameter of a metalens to improve the quality of optical imaging**

To further confirm our proof-of-concept devices are practical for imaging applications, we fabricated another metalens based on a 245-nm-thick boron nitride. The design parameter (i.e. the range of implemented phase) is the same as that for Device A shown in the main text, but this newly fabricated metalens is 120 µm in diameter, twice the diameter of Device A. Two boron nitride metalenses with different diameters were then used to create the image of the letter W for comparison. As shown in Fig. S7, the larger aperture metalens provides an improved intensity contrast with a lower exposure time due to its greater collection area, demonstrating improved image quality and enabling faster acquisition time.

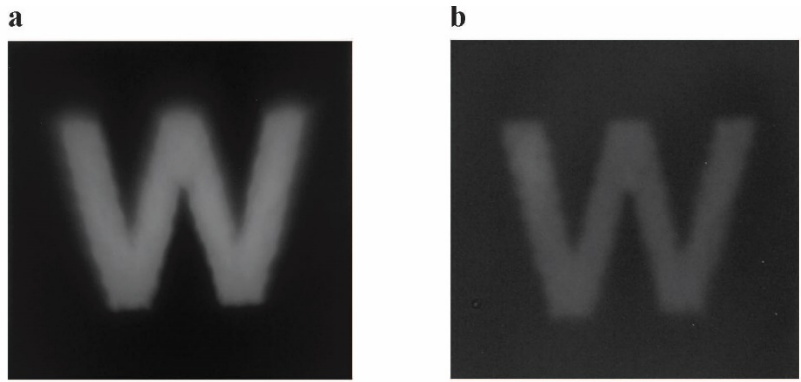

**Figure S7 Improving the quality of optical imaging. a.** The image was created by a 120-µm-diameter boron nitride metalens. The integration time for the CCD camera is 0.5 s. **b.** The image was created by a 60-µm-diameter boron nitride metalens (Device A, main text). The integration time for the CCD camera is 2s.